# Plausible Erklärungshinweise gegen die Überlichtgeschwindigkeit [20]

**Petra Schulz**
D-38116 Braunschweig, Theodor-Francke-Weg 65, Deutschland

**Kurzfassung**
Die vermeintlichen Überlichtgeschwindigkeiten werden vor allem durch Wechselwirkung der Materie mit der elektromagnetischen Strahlung aus dem Sender bzw. aus der eigenen Umgebung verursacht. Es werden unter geeigneten Bedingungen vermehrt anti-Stokessche Frequenzen emittiert, die zur Folge haben, daß die Brechzahl kleiner als 1 oder bei starker Laseranregung sogar negativ wird. Gleichzeitig wird durch die Wechselwirkung innerhalb der Materie die Lichtausbreitung gebremst. Bei den Versuchen mit der Lecher-Leitung und den Tunnel-experimenten ist das gemessene Signal überwiegend den schnellsten Weg über den metallischen Leiter gegangen. Die Lichtgeschwindigkeit bleibt eine Grenzgeschwindigkeit, die nicht überschritten werden kann. Die Brechzahl wird neu definiert und eine einfache Stoßtheorie für Stöße zwischen Photon und Teilchen vorgestellt. Durch die Relativbewegung zwischen Materie und Photon entscheidet die Materie oder das Photon darüber, ob es ein Neutrino ist oder nicht.

**Abstract**
The supposed supraluminal velocities of light are caused above all by interaction of matter with electro-magnetic radiation from the sender and/or the own environment. Anti-Stokes frequencies which result in the consequence that the refractive index becomes less than 1 or negative during strong optical laser stimulation even are emitted in a increased manner under suitable conditions. The light propagation is retarded simultaneously by interaction within the matter. In the case of the experiments with the Lecher wires and the tunnelling experiments mainly went the measured signal the fastest way over the metallic conductor. The velocity of light remains a limiting speed which can not be exceeded. The refractive index is defined in a new manner and a simple collision theory presented for impacts between photon and corpuscle. By the relative movement between matter and photon, the matter or the photon decides whether it is a neutrino or not.

**1. Einleitung**

In letzter Zeit (vor allem im letzten Jahr) ist in der wissenschaftlichen Literatur und in den Medien häufig etwas zum Thema Überlichtgeschwindigkeit veröffentlicht worden. Die vermeintlichen Lichtgeschwindigkeiten beflügeln die Phantasie einiger Physiker und lassen die Gehirnaktivität der Wissenschaftsjournalisten entflammen. (Beinah gezähmt klingt noch der Artikel des Journalisten RIPOTA [7]). Es tauchen vor allem in populären Veröffentlichungen nicht akzeptable Reizworte auf wie

- akausales Verhalten und
- Zeitreisen in die eigene Vergangenheit.

Der menschliche Verstand wird ganz schön in die Irre geleitet dank der herrschenden abstrakten Wellenmechanik und versperrt so den Denkweg zu einfacheren Modellvorstellungen. Die Physiker haben es ziemlich schwer, weil sie inzwischen so viele andere Geschwindigkeiten erfunden haben, darunter die Gruppen-, Phasengeschwindigkeit und Signalgeschwindigkeit. Das sind fast alles Größen, die sich physikalisch nicht messen lassen, und Anlaß zu Verwechslung geben. Z. B. wird die Phasengeschwindigkeit nur allzu oft mit normaler Lichtgeschwindigkeit verwechselt. Schuld daran mag sein, daß die Brechzahl (früher Brechungsindex) unvollständig definiert ist, unschönerweise als komplexe Größe. Brauchbar definiert ist die Brechzahl nur für den kleinen Bereich der geometrischen Optik.

Ich werde jedenfalls im folgenden nicht durch die wellenmechanische Brille gucken. Die ist zu unscharf für mein Thema, und außerdem kriege ich Bauchschmerzen. Durch meine Sichtweise möchte ich neue Impulse in die Physik hineinbringen. Dabei werde ich die endlich fällige Neudefinition der Brechzahl liefern.

**2. Versuche mit Überlichtgeschwindigkeiten**

Zunächst, wo kommen Versuche mit Überlichtgeschwindigkeiten vor? Hier wird eine Auswahl zusammengestellt:

Tesla-Experiment: Ein uraltes Experiment hat TESLA 1905 in einer Patentschrift beschrieben [9], das sein Nachfahre MEYL im Jahr 2000 nachgebaut hat [4]. Zwischen Sender und Empfänger wurde empfängerseitig eine Frequenzerhöhung im Bereich einiger Hertz festgestellt. Die Frequenz ließ sich sogar noch steigern, indem der Empfänger weiter vom Sender entfernt wurde. Beide Autoren nehmen in der Gleichung für die Frequenz $v$

$$v = c / \lambda \qquad \text{(Gleichung 1)}$$

mit $c$ = Vakuumlichtgeschwindigkeit die Wellenlänge $\lambda$ als konstant an. Aus den Verhältnissen der Frequenzen von Empfänger zum Sender wurde auf 1,5-fache Lichtgeschwindigkeit geschlossen. Das Voraussetzen einer konstanten Wellenlänge ist natürlich falsch. Auf das Problem der Frequenzerhöhung komme ich im Verlauf meines Vortrags noch zu sprechen.

Tunnel-Experimente: NIMTZ behauptete 1993, in einem Tunnel, einem verengten Rechteck-Hohlleiter, Mikrowellen mit 2,5-facher Lichtgeschwindigkeit beobachtet zu haben [6]. Spätere Versuche mit längeren Tunnelstrecken führten bei 8,7 GHz zu 4,7-facher Lichtgeschwindigkeit [2]. Hierbei wurde Mozarts Sinfonie Nr. 40 (verzerrt) übertragen, dieses Experiment wurde besonders populär.

Experimente mit der Lecher-Leitung: H. MÜLLER berichtete 2000 über sowjetische Versuche aus dem Jahre 1986 an einer Lecher-Leitung mit Dielektrikum [5]. Der Effekt wuchs mit der Länge der Lecher-Leitung. Die Forscher mischten zur stehenden Hochfrequenzwelle ein niederfrequentes Signal und stellten bei einer 15 km langen Leitung 20-fache Lichtgeschwindigkeit fest.

Brechzahl und anomale Dispersion: In der Literatur wird von Brechzahlen kleiner als 1 berichtet:

a) WOOD [11] beschreibt seine Experimente mit Natrium-Dampf aus dem Jahre 1904, für den er als minimale Brechzahl = 0,6 feststellte.

b) Im Lexikon der Physik aus dem Jahr 1998 wird erwähnt, daß Glas bei Röntgenbestrahlung Brechzahlen kleiner als 1 zeigt [1]. Was es aber mit den Brechzahlen kleiner als 1 auf sich hat, darüber schweigen sich Lehrbücher und Lexika aus.

Im Bereich der geometrischen Optik ist die Lichtgeschwindigkeit in Luft praktisch gleich der Lichtgeschwindigkeit im Vakuum. Somit kann die Brechzahl $n$ als das Verhältnis der Lichtgeschwindigkeit in Luft $c_{Luft}$ zur Lichtgeschwindigkeit im Medium $c_m$ angesehen werden.

$$n = c_{Luft} / c_m \qquad \text{(Gleichung 2)}$$

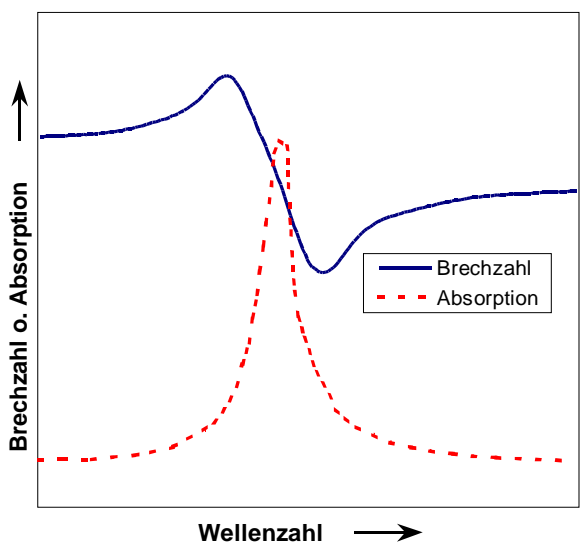

Abb. 1: Schematischer Verlauf der Brechzahl und der Absorption bei der anomalen Dispersion

Das ist sehr praktisch, weil man die meisten Versuche in Luft macht und man sich nicht unnütz ins Vakuum begeben muß. Die Brechzahlen der Stoffe, die in der geometrischen Optik verwendet werden, sind größer als 1. Die benutzten Materialien sind durchlässig und absorbieren fast kein Licht.

Bei stärkerer Absorption oder gegebenenfalls auch bei Emission (beides sind Fälle mit anomaler Dispersion) ist die Benutzung von Gleichung 2 nicht mehr zulässig (siehe z. B. Physik-Lexikon [1]). Die Neuinterpretation der Brechzahl als Quotient zweier Phasengeschwindigkeiten analog Gleichung 2 ist ein gefährlicher Weg, da er zu Verwechslungen führen kann. Diese Methode sollte den mathematisch denkenden Physikern vorbehalten sein und nicht zum Lehrstoff an Schulen gemacht werden.

c) WANG und Mitarbeiter [10] arbeiteten im Jahr 2000 mit Cäsiumgas im Bereich anomaler Dispersion mit Laserstrahlen und Magnetfeld. Durch Zurückrechnen folgerten sie bei ihrem Versuch auf eine -310-fache Lichtgeschwindigkeit bzw. eine Brechzahl von –310. In Abb. 1 wird schematisch der Verlauf der Brechzahl und der Absorption mit der Wellenzahl bei anomaler Dispersion wiedergegeben, der allerdings nicht für das WANG-Experiment [10] zutrifft. Die Wellenzahl ist Frequenz durch $c$ und dadurch gleichzeitig der Kehrwert der Wellenlänge. Ich habe hier die Wellenzahl anstatt der Frequenz aufgetragen, weil ich es von der Molekülspektroskopie her gewöhnt bin.

Für die Fälle, bei denen Überlichtgeschwindigkeit postuliert wurde, empfehle ich dringend, grundsätzlich die Wechselwirkung der elektromagnetischen Strahlung mit der Materie zwischen Sender und Empfänger maßgeblich in Betracht zu ziehen. Diese Wechselwirkung kommt zustande durch Zusammenstoß von Teilchen zwischen Sender und Empfänger mit Photonen des Senders (und gegebenenfalls auch mit Photonen aus der eigenen Umgebung). Materie wird bei der Wechselwirkung mit elektromagnetischer Strahlung enorm gesprächig und gleichzeitig lästig. Die Gesprächigkeit erkennt man an der Vielzahl der Linien im Spektrum. Jeder Peak entspricht quasi einem gesprochenem Wort. Die Lästigkeit ergibt sich aus der Tatsache, daß die Teilchen das Licht bremsen können. Hierauf gehe ich später noch ein. Beide Eigenschaften will ich unter dem spektroskopischen Effekt verstehen.

Wie bei meinem letzten DPG-Vortrag in Dresden über die Definition von Masse, Ladung und Spin [8] werde ich auch diesmal hauptsächlich wieder durch die spektroskopische Brille gucken. Neben dem spektroskopischen tritt bei den Tunnel-Experimenten noch ein zusätzlicher Effekt geringeren Ausmaßes auf, den ich Blenden-Effekt nenen möchte. Den will ich als erstes kurz erläutern, und zwar in der klassischen Darstellungsweise ohne Quantenmechanik. Damit befinde ich mich gleichzeitig auf dem Boden der geometrischen Optik.

## 3. Blenden-Effekt

Weshalb bewegt sich Licht im Tunnel schneller als in einem Rohr mit weiterem Durchmesser? Wir beobachten grundsätzlich, daß zwei gleichzeitig gestartete Lichtstrahlen den Detektor i. a. unterschiedlich schnell treffen, weil sie unterschiedlich lange Lichtwege zurücklegen. Der Mittelstrahl sowie alle zu ihm parallelen Strahlen kommen eher an als die Randstrahlen (s. Abb. 2).

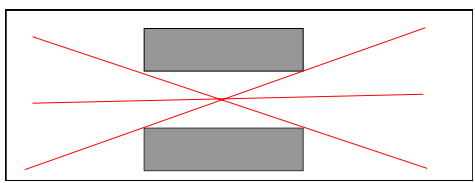

Abb. 2: Der breite Tunnel läßt alle Strahlen durch, eingezeichnet sind nur die unreflektierten.

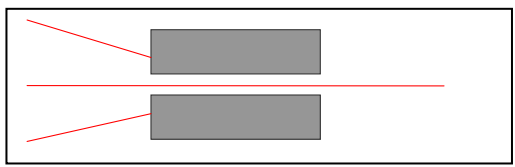

Abb. 3: Der schmale Tunnel unterdrückt Randstrahlen.

Im Tunnel werden Randstrahlen unterdrückt, s. Abb. 3. Somit ist dank des Blenden-Effektes durch den engen Tunnelquerschnitt der Anteil derjenigen Lichtstrahlen größer, die kürzere Lichtwege durchfliegen. Sofern den Lichtstrahlen eine zusätzliche Information wie die Sinfoniemodulation mitgegeben wurde, so muß diese Information schneller ankommen als bei einem Durchgang der Lichtstrahlen durch einen weiten Tunnel. Die schnellstmögliche Übertragungsgeschwindigkeit wird natürlich die Lichtgeschwindigkeit sein. Auf den Haupteffekt des Tunnelversuchs komme ich erst in Kapitel 5 zurück.

## 4. Spektroskopischer Effekt

Als ich über die anomale Dispersion sprach, stand ich bereits kurz vor der Erläuterung des spektroskopischen Effektes. Doch um ganz zu diesem Thema zu gelangen, muß ich trotzdem noch ein wenig ausholen. Nehmen wir zur Einstimmung die Clausius-Mosotti-Gleichung als Denkgrundlage an, die in etlichen Lehrbüchern neuerdings einfach unterschlagen wird:

$$\frac{\varepsilon - 1}{\varepsilon + 2} = const \cdot \rho \cdot (\alpha + \frac{\mu^2}{3kT}) \quad \text{(Gleichung 3)}$$

mit $\varepsilon$ = Dielektrizitätszahl (Permittivitätszahl), $\alpha$ = Polarisierbarkeit, $\mu$ = Dipolmoment, $k$ = Boltzmannkonstante, $T$ = absolute Temperatur, $\rho$ = Dichte des durchstrahlten Mediums. Ersetzen wir nach der Maxwell-Beziehung (Gleichung 4) $\varepsilon$ durch die Brechzahl $n$,

$$\varepsilon = n^2 \quad \text{(Gleichung 4)}$$

so erhalten wir

$$\frac{n^2 - 1}{n^2 + 2} = const \cdot \rho \cdot (\alpha + \frac{\mu^2}{3kT}) \quad \text{(Gleichung 3a)}$$

Der Einfachheit halber beschränke ich mich in meinen Ausführungen überwiegend auf ein Gas. Wenn Moleküle mit Licht beschossen werden, kann es sein, das sie das Licht sich einverleiben (es absorbieren), d. h., sie werden angeregt. Und später werden sie das Licht wieder ausspucken (emittieren). Wenn die Moleküle dabei ihre Polarisierbarkeit (ihren räumlichen Platzbedarf) ändern, sie werden verformt, dann wird man ein Ramanspektrum beobachten. Darauf werde ich im folgenden noch zu sprechen kommen.

Wenn sich durch den Beschuß mit elektromagnetischer Strahlung das Dipolmoment ändert, so wird sich ein IR-, Mikrowellen- oder sonstiges Spektrum zeigen. Das kann sowohl ein Absorptions- als auch ein Emissionsspektrum sein. Infolge der Wechselwirkung der Materie mit Licht wird sich also die Brechzahl als eine Funktion der eingestrahlten Frequenz oder Wellenzahl erweisen. Diese Aussage steckt nur indirekt (für den Wissenden) in der modifizierten Clausius-Mosotti-Gleichung 3a.

Der Raman-Effekt tritt auf, wenn Teilchen mit monochromatischen Photonen beschossen werden. Dann sind drei Fälle zu unterscheiden: 1. Das Teilchen reagiert nicht mit dem Photon (dieses Photon gibt sich im Spektrum als unverschobene Rayleigh-Linie zu erkennen). 2. das Teilchen absorbiert Licht. 3. Oder es emittiert den Überschuß eigener Energie, und zwar dasjenige Photon, das in der Vergangenheit vom Teilchen gefangen wurde.

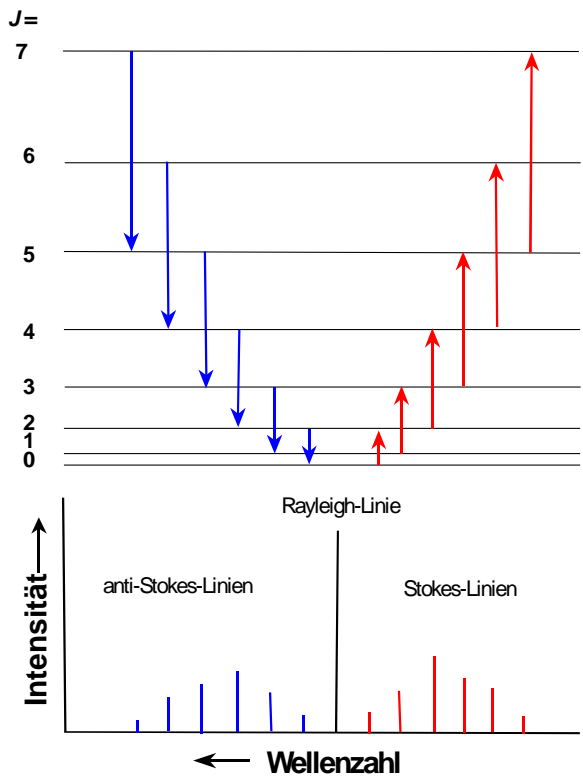

Abb. 4: Übergänge und Rotations-Raman-Spektrum eines zweiatomigen Moleküls. Die Stokes-Linien kommen durch Absorption und die anti-Stokes-Linien durch Emission zustande. (Die Gesamtdrehimpulsquantenzahl $J$ ändert sich dabei um zwei Einheiten.)

In Abb. 4 ist das beinahe einfachste Rotations-Raman-Spektrum dargestellt, und zwar das von einem zweiatomigen Molekül, zunächst nur die schematische Darstellung, In der linken Hälfte neben der eingestrahlten Rayleigh-Linie entstehen durch Emission die anti-Stokes-Linien und in der rechten Hälfte durch Absorption die Stokes-Linien. Die anti-Stokes-Linien weisen eine höhere Wellenzahl auf als die Rayleigh-Linie.

In Abb. 5 ist nun ein experimentelles Rotations-Raman-Spektrum von $^{15}N_2$ dargestellt. (Ich habe leider kein hochaufgelöstes Spektrum von Luftstickstoff $^{14}N_2$ finden können.) Wenn man mit niedrigen Intensitäten zufrieden ist, bräuchte man eigentlich gar keine erregende Rayleigh-Einstrahlung. Denn die energiereicheren Stickstoffmoleküle werden ihr gespeichertes "Licht" und damit ihre Rotation abgeben, bevor sie vom Sendestrahl getroffen werden, weil sie nur von Nachbarteilchen gestoßen werden. Das ist nichts besonderes, da die Rotation von Gasmolekülen bereits bei Zimmertemperatur angeregt ist. Andere, weniger angeregte Moleküle werden dagegen erst mal elektromagnetische Strahlung aufnehmen und dadurch ihren Rotationszustand erhöhen. Damit der Meßeffekt deutlich wird oder weil man zufällig ein Ramangerät mit entsprechend energiereichen Lasern besitzt, mischt man die Materie durch die Rayleigh-Einstrahlung anheizend auf.

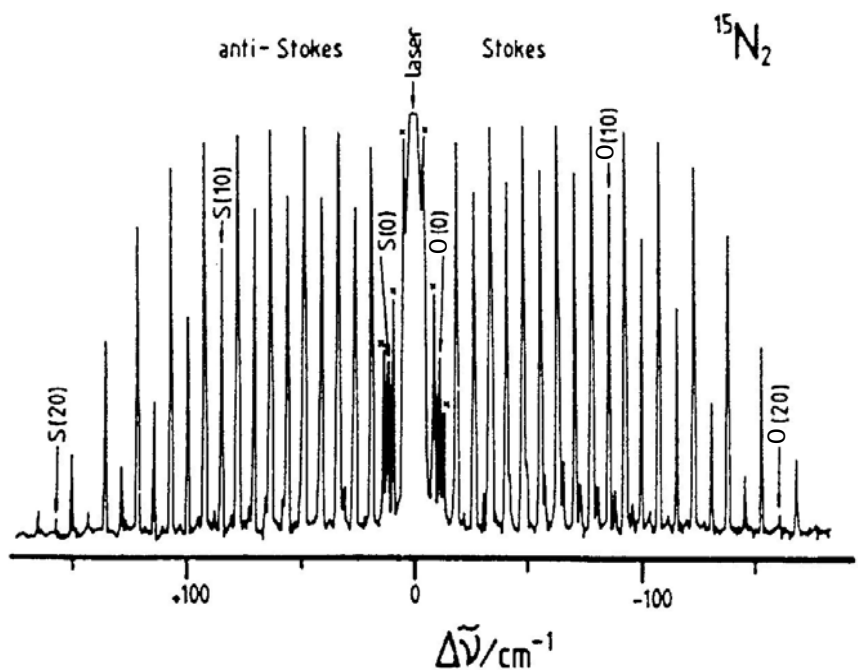

Abb. 5: Rotations-Raman-Spektrum von $^{15}N_2$ (aus HOLLAS [3]). Die mit einem Kreuz markierten Linien nahe der Laserlinie gehören nicht zum Spektrum.

## 5. Zur Deutung der vermeintlichen supraluminalen Experimente

### 5.1 Die Wechselwirkung

Sehr viele Versuche mit vermeintlichen Überlichtgeschwindigkeiten lassen sich wohl erklären, wenn man annimmt, daß beim Anschalten des Senders die Materie einen Raman-Effekt zeigt. Sie emittiert dabei im anti-Stokeschen Bereich Strahlung höherer Frequenz, als der Sender abgibt. Der Effekt fällt um so drastischer aus, je mehr der anti-Stokessche Teil des Spektrums angeregt werden kann.

Wer waren die angeregten Teilchen im Empfänger? Die angeregten Teilchen bei den Tesla- und Nimtz-Versuchen dürften Luftmoleküle gewesen sein, beim Wang-Experiment das Valenzelektron der Cäsiumatome, bei den russischen Experimenten das nicht näher beschriebene piezoelektrische Dielektrikum an der Lecher-Leitung.

Der Haupterreger war sicher meistens der Sendestrahl. Welche Erregungsarten spielten zusätzlich eine Rolle? Die Anregung der Luftmoleküle erfolgte durch die Umgebungstemperaturen und die Sendefrequenz, die Anregung der Cäsiumatome durch ein schwaches Temperaturbad von 30 °C, ein schwaches Magnetfeld und vor allem durch mehrfache Laserbestrahlung.

Der negative Brechungsindex beim Wang-Experiment kommt durch eine resultierende Emission zustande, die den Absorptionsprozeß während der Messung deutlich übersteigt.

### 5.2 Der Wettlauf durch den Leiter und den Nichtleiter

Für die beschriebene Deutung der Experimente mit vermeintlichen Überlichtgeschwindigkeiten als Raman-Effekt spricht ferner die Tatsache, daß bei dem Tesla-Experiment sowie bei den Versuchen mit der Lecher-Leitung und den Tunnelexperimenten die Länge der Versuchsstrecke einen Einfluß auf das Meßergebnis hatte.

Eine größere Durchstrahlungslänge bedeutet eine größere Wahrscheinlichkeit für die Photonen mit Molekülen des Dielektrikums (einschließlich der Moleküle untereinander durch Photonenaustausch) wechselwirken zu können. Der quantitative Zusammenhang zwischen Absorption $a$ durch der Durchstrahlungslänge $d$ wird durch das Lambert-Beersche Gesetz wiedergegeben:

$$a = \varepsilon \cdot x \cdot d \qquad \text{(Gleichung 5)}.$$

$\varepsilon$ ist der molare (dekadische) Absorptionskoeffizient oder auch der dekadische Extinktionskoeffizient oder die dekadische Absorption (dekadisch bedeutet bezogen auf die Basis 10 anstelle zur Basis der Eulerzahl $e$). $x$ ist hier die Konzentration des durchstrahlten Stoffes.

Die Wechselwirkung des Lichts mit der Materie durch Absorption und Re-emission beansprucht allerdings zusätzliche Zeit und bremst damit die Lichtübertragung. Die Teilchen geben das einmal gefangene Licht nicht so schnell wieder her. Sie gehen mit dem Licht nämlich erst noch spazieren. Und Teilchen sind nun mal nicht so schnell wie das Licht.

Somit muß für die Versuche mit der Lecher-Leitung und den Tunnelexperimenten noch folgendes angenommen werden: Das gemessene Signal wird den widerstandsärmsten Weg eingeschlagen haben, und zwar den durch den metallischen Leiter, bevorzugt wohl durch die Schicht nahe der Metalloberfläche. Dieses Signal war schneller als das Vergleichssignal durch Luft (bei NIMTZ) bzw. durch das Dielektrikum (s. H. MÜLLER).

| Beobachteter Effekt | Experiment | Grund |
|---|---|---|
| Frequenzerhöhung | TESLA/ MEYL | anti-Stokes-Linien emittiert |
| Emissionsspektrum Brechzahlspektrum Brechzahl negativ | WANG | sehr starke anti-Stokes-Emission |
| Lichtübertragung verzögert → schnellerer Weg über metallischen Leiter | Lecherleitung + NIMTZ | starke Absorption + gleichstarke (?) anti-Stokes-Emission |

Tab. 1: Versuch einer Deutung der Experimente mit vermeintlichen Überlichtgeschwindigkeiten

In Tabelle 1 sind die wahrscheinlichen Deutungsmöglichkeiten für die Versuche mit vermeintlichen Überlichtgeschwindigkeiten zusammengestellt.

## 6. Neudefinition der Brechzahl

Ein wichtiges Nebenprodukt dieser Arbeit ist die Neudefinition der Brechzahl: **Die Brechzahl ist im wesentlichen die 1. Ableitung der Absorption $a$ nach der eingestrahlten Frequenz**.

$$n = 1 + const \cdot \frac{da}{d\nu} \quad \text{(Gleichung 6)}$$

Da die Frequenz $\nu$ in Hz gemessen wird, hat die Konstante die Größeneinheit Hz.

$$const = \ln(10) \text{ Hz}$$

Gleichung 6 ist allgemeingültig und nicht auf den kleinen Bereich der geometrischen Optik beschränkt.

Für die Absorption existieren die Synonyme Extinktion, Absorbanz und optische Dichte [18]. Damit es keine Mißverständnisse gibt, was ich unter Absorption verstehe:

$$a = \lg(I_0 / I) \quad \text{(Gleichung 7)}$$

Absorption ist der dekadische Logarithmus vom Verhältnis der Intensität des einfallenden Lichts $I_0$ zur Intensität des austretenden Lichts $I$. Und Intensität bedeutet Energie $E$ durch das Produkt von Zeit $t$ und Fläche $F$.

$$I = \frac{E}{t \cdot F} \quad \text{(Gleichung 8)}$$

Bei Emission ist $I$ größer als $I_0$. Damit ist das Verhältnis $I_0/I < 1$ und sein Logarithmus halt negativ (siehe Gl. 7).

Die Brechzahl ist reell. Saubere (schmale) Frequenzen liefern in der Praxis genaue Brechzahlen. Es müssen Brechzahl- oder/und Absorptionsspektren bzw. Emissionsspektren für verschiedene Auflösungen archiviert werden.

Für die Elektrizitätslehre kann als Dielektrizitätszahl $\varepsilon$ unmittelbar die Maxwell-Beziehung $\varepsilon = n^2$ (Gleichung 4) angewendet werden. Wie $n$ ist auch $\varepsilon$ eine reelle Größe.

Wird Materie mit Licht bestrahlt, kann man, wie aus der Gleichung 6 ersichtlich, drei Grenzfälle zu unterscheiden
- $n > 1$ bei der Absorption
- $n = 1$ bei der Durchlässigkeit z. B. durch das Vakuum oder durch Luft
- $n < 1$ bzw. $n < 0$ bei der Emission.

Diese Fälle werde ich im nächsten Kapitel etwas ausführlicher an Hand von wichtigen Stoßprozessen beleuchten.

### 7. Photonenstöße mit Teilchen

In den Abbildungen 6 bis 10 werden wichtige Grenzfälle von Photonenstößen mit Teilchen dargestellt. Das sind alles Prozesse, die in einer stehenden Welle ablaufen können. Die in der Tabelle 1 zusammengestellten Versuche wurden mit stehenden Wellen durchgeführt. In den Abbildungen 6 bis 10 steht links der Zustand vor dem Stoß und rechts der Zustand nach dem Stoß. In Abb. 6 und 7 trifft das Photon auf ein ruhendes Teilchen und in Abb. 8 bis 10 auf ein entgegengesetzt fliegendes Teilchen.

Das reine Photon (das Teilchen, das mit $c$ durchs Vakuum fliegt) ist als Pfeil gezeichnet, das ruhende Teilchen als Kreis, das angeregte Teilchen als Linearkombination von Teilchen und Photon. Für diesen Zwitterzustand zwischen Teilchen und Photon hat sich neuerdings der Begriff Polariton eingebürgert [19]. Ein Ruhmassenzuwachs beim bewegten Teilchen durch einen Bremsprozeß wird durch einen dickeren Kreis symbolisiert. Die Größe der Frequenz wird durch die Länge des Pfeils dargestellt. Als Photon verstehe ich ein Teilchen, das im einfachsten Fall transversal (zwei Richtungen quer zur Flugrichtung) oder/und longitudinal polarisiert ist (jeweils in bzw. entgegengesetzt zur Flugrichtung). Für die Flugrichtung gibt es zwei Zustände: hin und zurück. Energetische Überlegungen werden vom Standpunkt des Teilchens gemacht.

### 7.1 Absorption

In Abb. 6 haben Teilchen und Photon die gleiche Energie, deshalb kann das Photon vom Teilchen absorbiert werden. Das ist ein maximal inelastischer Stoß (maximaler Compton-Effekt). Das Teilchen benimmt sich wie ein schwarzes Loch. Absorption verursacht eine Dämpfung. Das Schwarze Loch wird bewegt. Anders ausgedrückt: aus dem Teilchen wurde ein Polariton.

Ob das in einem Mehrteilchensystem wirklich ein inelastischer Prozeß war, wird sich erst zeigen, welche anderen Stoßpartner noch vorhanden sind und wirksam werden. Bei gleichartigen Teilchen wie Elektronen mit gleichem Energieinhalt oder Billardkugeln wird es sich im Vielteilchensystem um einen elastischen Stoß handeln (beinahe elastisch z. B. bei der elektrischen Leitung durch ein Metall bei Zimmertemperatur).

Den Absorptionsprozeß kann ich auch anders deuten. Das Photon wurde vom Teilchen gekidnappt. Dem Teilchen wurde Schmerz zugefügt. Der Frequenzübertrag durch das Photon ist ein Maß für den Schmerzzuwachs im Teilchen.

Wenn das Teilchen mit dem eingefangenen Photon reden könnte, würde es sagen: "Ich muß gehn".

### 7.2 Transparenz und das Neutrino

In Abb. 7 haben Teilchen und Photon unterschiedliche Energien oder nicht ineinander paßgerechte Raum-Zeit-Strukturen. Diese Formulierung klingt scheußlich, ich weiß. Was ich damit meine, erkläre ich an einem Beispiel:

Stellen wir uns ein schwingendes Etwas als ein Teilchen vor, das so klein ist wie ein Elektron, Positron oder das neutrale Verschmelzungsprodukt aus Elektron und Positron. Hier nur eine Ergänzung zu [8]: Das Elektron ist übrigens eine spinnende Kugel, das Positron die gegensinnig spinnende Kugel. Ein denkbares hochsymmetrisches Verschmelzungsprodukt aus Elektron und Positron, das ich vor mir sehe, das ist die atmende Kugel. (Die atmende Kugel kommt zustande, wenn Elektron und Positron gleichzeitig am selben Ort senkrecht aufeinander stehen.) Ich will aber das schwingende Teilchen nicht weiter spezifizieren und einfach Parton nennen als das potentielle Wechselwirkungsteilchen für elektromagnetische Strahlung. Ich könnte auch ein größeres Teilchen gewählt haben, etwa ein Atom, Molekül oder Komplizierteres. Aber dann müßte das potentiell wechselwirkungsfähige Photon genauso strukturiert sein wie die genannten Teilchen selbst.

Angenommen, das Parton schwingt senkrecht zur Flugrichtung des reinen Photons. Die Ruhelage des Partons soll sich nicht verändern. Wenn das Parton sich gerade vom Schwingungsmittelpunkt entfernt hat und das Photon diesen Punkt durchfliegt, kann es keine Wechselwirkung zwischen Teilchen und Photon geben. Somit geht das Photon durch das Teilchen hindurch. Das Teilchen ist also durchsichtig. Ich habe damit leidlich das Kapitel Polarisation angesprochen. Teilchen und Photon haben in dem geschilderten Beispiel also den ausgezeichneten Phasenunterschied, daß sie sich nie gleichzeitig treffen können.

In diesem Spezialfall ist das Photon für das Teilchen ein Neutrino. Es findet keine Dämpfung statt. **Photon und Neutrino sind identisch. Durch die Relativbewegung zwischen Materie und Photon entscheidet die Materie oder das Photon** (falls es energiereicher ist als das Materieteilchen) **darüber, ob es ein Neutrino ist oder nicht.** Schließlich ist das Neutrino keinesfalls nur ein elitäres Produkt der Hochenergiephysik! Das Photon einer Frequenz kann für den einen Stoff ein Photon und für den anderen ein Neutrino sein. Verwendet man mit dem gleichen Stoff unterschiedliche Einstrahlungsrichtungen (um 90 ° innerhalb der Papierebene gedreht sowie um 90 ° über die Papierebene gedreht), so kann man durchaus sogar zu drei unterschiedlichen Beobachtungsergebnissen kommen, denn nicht alle Stoffe sind isotrop). Die Brechzahl ist zwar reell, aber ein Tensor.

Mit den Vokabeln aus der Schwarzweiß-Photographie gesprochen: **Teilchen sind das Negativ von Photonen und umgekehrt.** Man kann das Ganze auch unter Jin-Jang-Theorie abheften. Wie groß die Ruhemasse des energieärmsten Neutrinos ist, habe ich in [8] bereits vorgestellt. Da keine Wechselwirkung stattgefunden hat, nichts absorbiert und nichts emittiert worden ist, ist gemäß Gleichung 6 die Brechzahl $n = 1$. Transparente Vorgänge sind schmerzlos.

Hier noch eine kurze Bemerkung zum Vielteilchensystem: Wenn in einem Metallgitter alle Atome im Gleichtakt schwingen (schunkeln), dann kann das auf das Metall geschossene gleichgetaktete kohärente Licht ungehindert hindurchfliegen. Das dürfte ein Zustand bei der Supraleitung sein.

Wenn aber der Phasenunterschied zwischen dem Teilchengitter und dem Photonengitter gerade so eingestellt ist, daß die Photonen die Teilchen im Schwingungsmittelpunkt treffen, dann haben wir den maximalen Leidenszustand erreicht, die totale Absorption. Das ist die Phononenschwingung, wie wir sie von der Ausbreitung der Schallwellen her kennen.

**7.3 Strahlungslose Reflexion oder die Lichtbremse**
In Abb. 8 trifft ein Photon auf ein angeregtes Teilchen, erzeugt durch Abb. 6, das ihm entgegenfliegt. Das reine Photon und das Photon im Teilchen haben die gleiche Energie und vereinigen sich. Ihre Wellen löschen sich aus. Es verschwindet kinetische Energie, sie wird in potentielle Energie umgewandelt, angedeutet durch die dickere Kreislinie. Das Teilchen gewinnt Ruhmassenzuwachs $m_0$. Es findet eine Bose-Einstein-Kondensation statt. Das Schwarze Loch wird schwerer. Der Massenzuwachs ergibt sich ergibt sich aus der Formel für den maximalen Energieübertrag beim Compton-Effekt.

$$h \cdot \nu = \tfrac{1}{2}\, m_0 \cdot c^2 \quad \text{(Gleichung 9)}$$

$h$ ist das Plancksche Wirkungsquant und $\nu$ der Frequenzübertrag. $c$ ist natürlich die Vakuumlichtgeschwindigkeit. Es findet eine Dämpfung statt. Das Teilchen absorbiert, deshalb ist $n = 1$. Es ist ein Dunkelfeld-Polariton entstanden.

Ich verstehe gar nicht, weshalb man Gleichung 9, die Compton-Gleichung, noch nicht metrologisch ausnutzt und sich immer noch mit dem schmutzigen Urkilogramm abmüht.

Die strahlungslose Reflexion ist geeignet, um den radioaktiven Zerfall innerhalb einer stehenden Welle auszulöschen[15]. Das auslöschende Licht muß von dem radioaktiven Stoff selbst stammen. Bisher wurde noch nicht die geeignete Frequenz angewendet. Jedes Isotop benötigt individuelle Frequenzen.

Die Totalbremsung ist - wie wir aus der Alltagserfahrung von Auffahrunfällen durch Geisterfahrer her wissen - ein schmerzhafter Prozeß. Ähnlich wie bei der Absorption ist der Ruhmassenzuwachs $m_0$ ein Maß für den Schmerzzuwachs für Photon und Teilchen. Der Bremsprozeß wirkt für einen außen stehenden ruhenden Beobachter, als würde das Licht auf der Stelle stehen, dabei ist es nur im Teilchen gefangen.

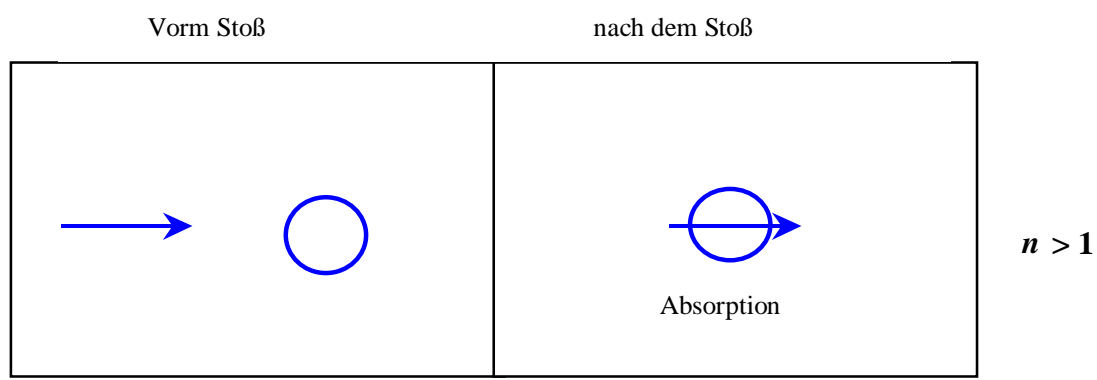

Abb. 6: Absorption eines Photons (maximaler Compton-Effekt). Das Schwarze Loch wird bewegt.

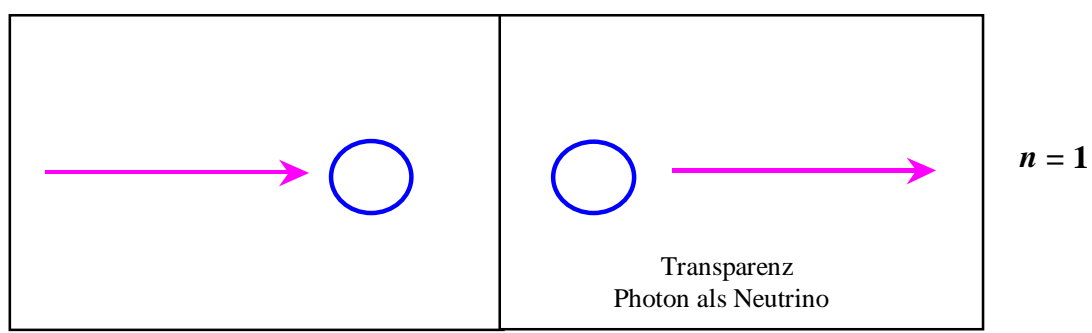

Abb. 7: Durchstrahlen eines Teilchens. Das Photon ist relativ zum Teilchen ein Neutrino.

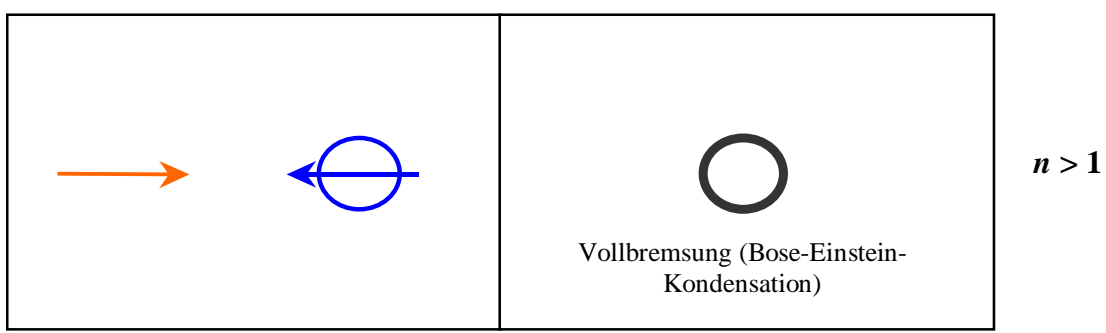

Abb. 8: Das Photon bremst das Teilchen total, das bewegte Schwarze Loch wird schwerer (Bose-Einstein-Kondensation).

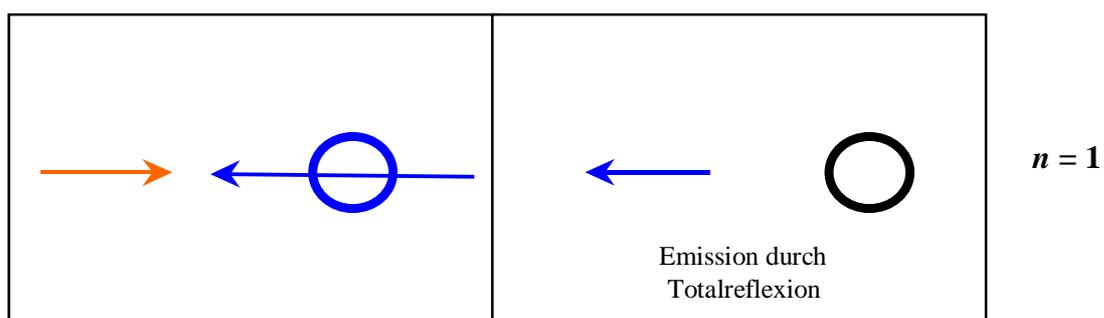

Abb. 9: Das photonenreichere Teilchen wird durch das reine Photon gebremst. Bei der Totalreflexion emittiert das getroffene Teilchen so viel, wie es absorbiert hat.

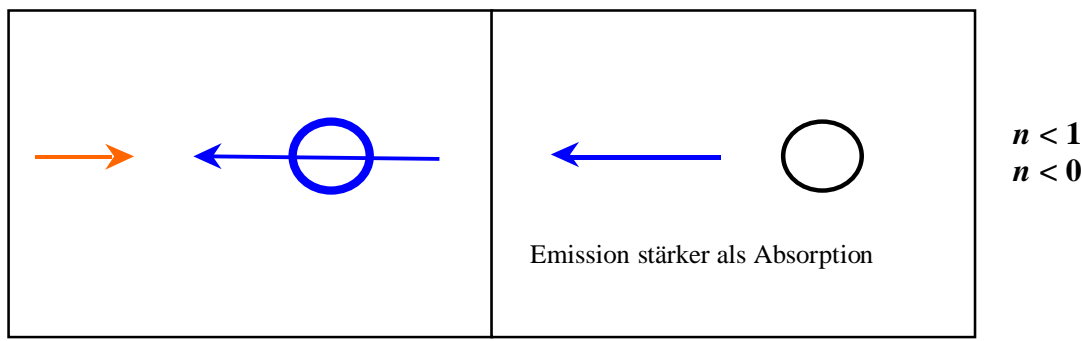

Abb. 10: Bei der Frequenzverdopplung ist die Emission doppelt so stark wie die Absorption.

### 7.4 Totalreflexion durch Emission/Spiegelung
In Abb. 9 fliegen das Photon (orange) und das angeregte Teilchen (blau) aufeinander zu. Das reine Photon hat eine geringere Energie als das Photon, das im Teilchen absorbiert ist. Die Frequenz des reinen Photons wird durch eine gleich große Frequenz aus dem Teilchenphoton neutralisiert (Wellenauslöschung wie im vorangegangenen Beispiel). Das noch überschüssige Photon im Teilchen hat eine verminderte Frequenz, sie verbleibt nicht im Teilchen. Die Neutralisation der entgegengesetzten Photonen führt zu einem Ruhmassezuwachs. Das noch übrig gebliebene halbe Photon wird emittiert. Insgesamt ist die Energiebilanz in Abb. 9 ausgeglichen. So viel Energie, wie das Teilchen absorbiert hat, hat es auch emittiert, also $n = 1$. Das emittierte Photon wurde symbolisch blau gemalt, um so die Herkunft vom ehemaligen Besitzer anzuzeigen.

**Die Totalreflexion kommt also durch Emission zustande.** In Abb. 9 findet im Photon des Teilchens eine Frequenzhalbierung statt. Die Totalreflexion ein schmerzhafter Prozeß, zusammengesetzt aus zwei einzelnen schmerzhaften Vorgängen.

### 7.4.1 Frequenzverdopplung
Schließlich ist in der Sammlung der exemplarischen Stoßprozesse zwischen Teilchen und Photon der Fall der Frequenzverdopplung in Abb. 10 abgebildet. Frequenzverdopplung tritt auf, wenn ein Photon und ein höher angeregtes Teilchen aufeinander zufliegen und das Photon im Teilchen eine dreimal so große Energie hat wie das reine Photon. Das Teilchen kommt zur Ruhe und emittiert ein Photon doppelter Frequenz im Vergleich zum reinen Photon. Das ist ein Vorgang, bei dem die Emission die Absorption übersteigt. Deshalb ist n < 1 bzw. bei starker Emission auch n < 0. Das ist dann ein Vorgang mit negativer Dämpfung, ein schmerzhafter Prozeß.

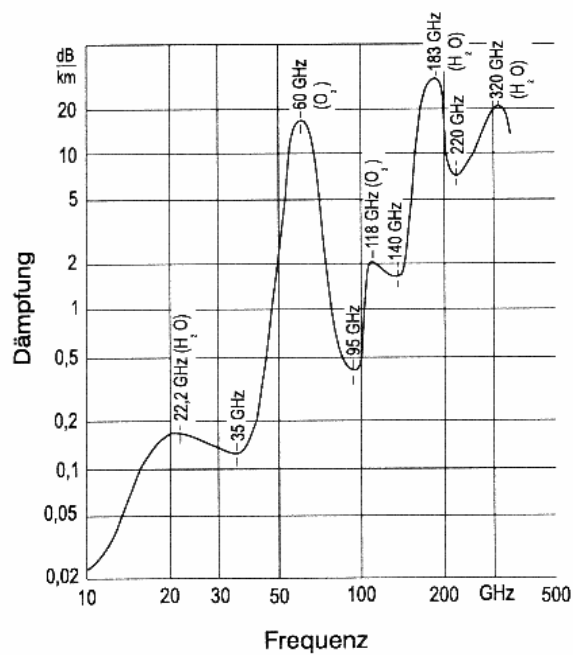

Abb. 11: Dämpfung von Mikrowellen durch Luft [12], [13]

### 7.4.2 Stoßprozesse und 2. Hauptsatz
Brechzahl < 1 bedeutet gleichzeitig einen Prozeß, bei dem nach dem 2. Hauptsatz (Gleichung 10)

$$\Delta G = \Delta H - T\Delta S \qquad \text{(Gleichung 10)}$$

die Freie Reaktionsenthalpie $\Delta G$ negativ ist. $\Delta H$ ist die Reaktionsenthalpie (Gesamtenergie), $T$ die absolute Temperatur und $\Delta S$ die Reaktionsentropie. $\Delta G$ ist die Energie des emittierten Photons. $T\Delta S$ stellt die neutralisierte Energie dar, also ein Maß für den Energiezuwachs durch den Bremsvorgang (siehe z. B. Abb. 8).

Einfacher ist der Fall bei der Durchsichtigkeit. Man steckt geordnete Energie in Form eines parallelen Photonenstroms ein und hofft, daß dieser am Ende wieder vollständig austritt. Das sind in der Wissenschaft, Natur und Technik wichtige Vorgänge, die freiwillig ablaufen (anschauliche Darstellung s. [17]).

**Chemie ist schließlich nichts anders als die Kunst, durch geeignete Anordnung von Teilchen, das in ihnen gefangene Licht zu befreien.**

### 8. Dämpfung in der Atmosphäre

Für Zweifler hier einen indirekten Beweis für den spektroskopischen Effekt in Luft. In Abb. 11 ist die Dämpfung von atmosphärischer Luft einer Schichtdicke von 1 km in Abhängigkeit von der Frequenz dargestellt, und zwar im Mikrowellenbereich von 10 bis rund 300 GHz [12]. Die Maxima werden den Wasser- ($H_2O$) und Sauerstoffmolekülen ($O_2$) zugeordnet [12], [13].

Der Vollständigkeit halber ist noch in Abb. 12 die atmosphärische Dämpfung von 10 GHz bis 1000 THz (also bis zum sichtbaren Bereich) wiedergegeben.

Die Dämpfung in Abb. 11 soll angeblich durch Schwingungsanregung hervorgerufen sein [13]. Ob das stimmt? Angesichts der Vielzahl an Möglichkeiten für eine Schwingung durch Grund-, Oberschwingung, Kombinationsschwingungen, Differenz- und heiße Banden mag eine sichere Zuordnung schwer fallen. (Heiße Bande bedeutet, daß ein bereits angeregtes Molekül nochmals angeregt wird.) Darüber hinaus ergeben sich auch zwischen den Molekülen verschiedener Stoffe Kombinations- und ähnliche Schwingungen. Was würden die Spektroskopiker konkret dazu sagen?

### 9. Luft und die Spektroskopie

Mit atmosphärischer Luft würden sich Spektroskopiker niemals befassen, das wäre einfach "Schweinkram": In der Luft befinden sich viel zu viele verschiedene Moleküle ($N_2$, $O_2$, $CO_2$, variable Mengen an $H_2O$), die alle Spektrallinien abliefern würden und vor allem - wie schon erwähnt - die Anzahl der möglichen Kombinations- und Differenzbanden unnütz erhöhen würden. Das Ganze würde zu allem Überfluß bei einem solchen abnormen hohen Druck wie Atmosphärendruck geschehen. Spektroskopiker bevorzugen nämlich reine Stoffe, am liebsten sogar isotopenrein (wenn es sowohl die Experimentierkunst als auch die Finanzen erlauben), denn nur so kann die Linienvielfalt entscheidend eingeschränkt werden. Ein Spektroskopiker würde niemals sein Gas bei 101,3 kPa beobachten wollen, weil die Linien so satt und breit werden und die Auflösung dabei verloren geht.

Wenn man ihn vielleicht doch bewegen könnte, ein Spektrum von Luft aufzunehmen (höchstens als Dienstleistung), dann müßte das bei reduziertem

Abb. 12: Dämpfung der Atmosphäre von 10 GHz bis 1000 THz (aus [14])

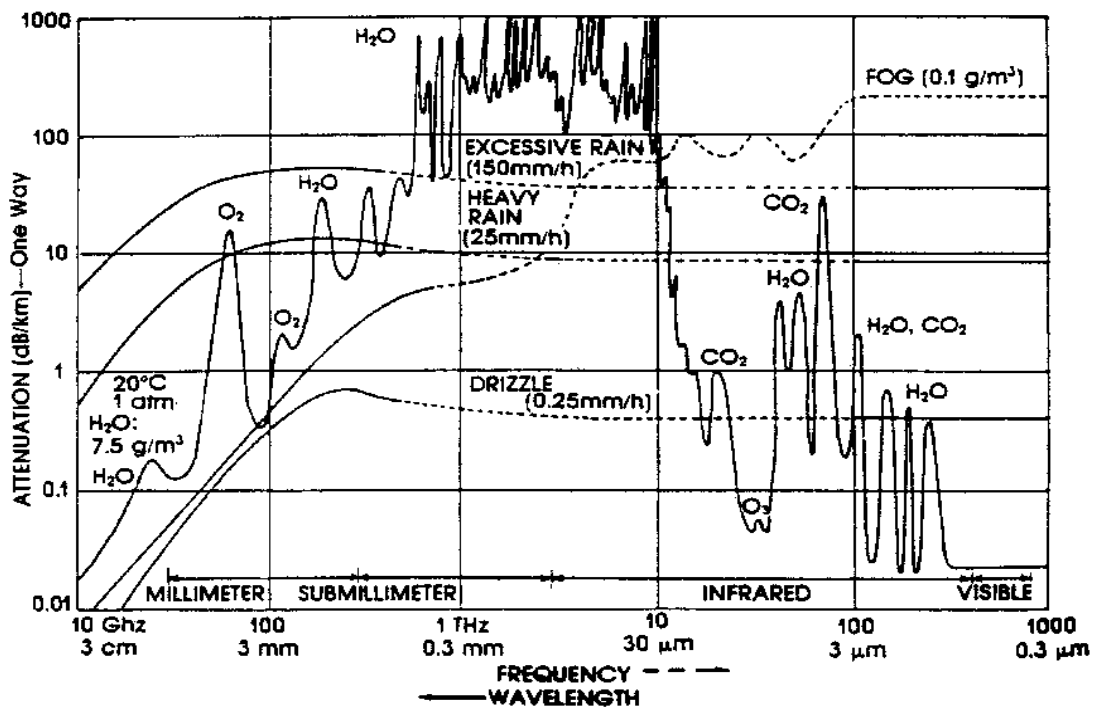

Druck geschehen. Der Vergleich des Spektrums der Luft mit den Spektren der einzelnen Luftbestandteile würde einige Diskrepanzen aufweisen. In Luft stoßen Moleküle des einen Stoffes auch auf Moleküle des anderen Stoffes. Dabei können die Molekülpeaks verschoben und gegebenfalls in der Intensität verändert werden. Die Spektroskopiker nennen solche Veränderungen Resonanz. Solche gemischten Resonanzen könnten eines Tages ein Motiv für die spektroskopisch forschenden Wissenschaftler sein, sich doch auch einmal mit Luft zu befassen.

Aber eine sinnvolle Weise, sich mit diesem Problem zu befassen, ist erst möglich, wenn im Bewußtsein der Spektroskopiker geschrieben steht, daß Spektren nur durch Stoßprozesse zustande kommen und nicht bloß aus der abstrakten Wellenfunktion hervorgehen.

### 10. Schluß

Nach den vorangegangenen Erläuterungen folgt nun der Schluß: **Die Lichtgeschwindigkeit bleibt eine Grenzgeschwindigkeit, die nicht überschritten werden kann.** Sicher kann man die Physik auch anders deuten, aber ich halte meine Darstellung für einfacher, ehrlicher und darum menschlicher.

### 11. Danksagung

Ich danke hier den folgenden Physikern, die mir als armem Chemiker kurz vor dem entsetzlich frühen Redaktionsschluß durch fetzende, aufklärende Gespräche, Emails oder/und Briefe auf die Sprünge geholfen haben: den Herrn Dr. G. Galeczki, Dr. P. Marquardt (beide aus Köln), Dr. Frank Stolpe aus Göttingen.

### 12. Literarisches [16]

Staubkorn
wer preist dich
wer streichelt dich
wer singt dir ein Liebeslied
Du, der du unaufhörlich
hin- und hergestoßen wirst
von den Moleküln der Luft
deinen jüngren Brüdern
die zischen und jagen
so daß du aufwärts schießt
dich winselnd wendest
abwärts sinkst
und als abermals Getretener
in die Höhe fliehst
Wer teilt
Dein Leid mit dir

**Nicht-ich**
Ich bin das Ich
das niemals ist
weil es in mir wühlt
dann aus mir dampft
und weil mich die Umwelt
stets beschießt
und mich nie in Ruh läßt
Was ich bin
Das ist das
Nicht-ich